\documentclass[aps,prl,reprint,groupedaddress,showpacs,amsmath,amssymb]{revtex4-1}

\usepackage{graphicx}
\usepackage{enumerate}
\usepackage{times}
\usepackage{natbib}

\DeclareMathAlphabet{\bi}{OML}{cmm}{b}{it}

\long\def\symbolfootnote[#1]#2{\begingroup
\def\thefootnote{\fnsymbol{footnote}}\footnote[#1]{#2}\endgroup}

\begin{document}

\title{The struggle for space: Viral extinction through competition for cells}

\author{Jos\'e A.~Cuesta$^1$, Jacobo Aguirre$^2$, Jos\'e A.~Capit\'an$^1$ and Susanna C.~Manrubia$^2$}

\affiliation{$^1${\it Grupo Interdisciplinar de Sistemas Complejos (GISC),
Departamento de Matem\'aticas, Universidad Carlos III de Madrid,
Legan\'es, Madrid, Spain}\\
$^2${Centro de Astrobiolog\'{\i}a, CSIC-INTA, Torrej\'on de Ardoz,
Madrid, Spain}}

\begin{abstract}
The design of protocols to suppress the 
propagation of viral infections is an enduring enterprise, especially hindered by limited
knowledge of the mechanisms through which extinction of infection propagation
comes about. We here report on a
mechanism causing extinction of a propagating infection due to intraspecific
competition to infect susceptible hosts. Beneficial mutations
allow the pathogen to increase the production of progeny, while
the host cell is allowed to develop defenses against infection. When the number
of susceptible cells is unlimited, a feedback runaway co-evolution
between host resistance and progeny production occurs. However, 
physical space limits the advantage that the virus can obtain from increasing offspring 
numbers, thus infection clearance may result from an increase in host
defenses beyond a finite threshold. Our results might be relevant to 
better understand propagation of viral infections in tissues with mobility 
constraints, and the implications that environments with different geometrical 
properties might have in devising control strategies. 
\end{abstract}

\pacs{87.23.-n, 87.10.-e, 87.18.-h}

\maketitle

%Cellular parasites are an unavoidable outcome of the very
%evolutionary process~\cite{Koo06}. 
One of the common strategies used by viruses 
to ensure fast adaptation is the steady production of mutants 
and of enormous amounts of progeny, two characteristics that
challenge the design of therapies able to induce viral extinction.
Increased mutagenesis has been successfully used
{\it in vitro} to eradicate infectivity, though there is no general
agreement on the pathways through which extinction
supervenes~\cite{Wil05,Bul05,Tak07,Man10}. 
In mutational meltdown, the population becomes extinct because the 
average growth rate falls below one, through
a process different from the error threshold predicted by classical
quasispecies models~\cite{Eig71} or 
from stochastic extinction of infectivity~\cite{Gra05,Ira09}.
The progress of an infection is further conditioned by the geometry of the space
where it occurs~\cite{Pet04,Bar08,Agu08}. 
Experiments with viruses infecting bacteria
have highlighted differences in the internal structure of 
populations evolving in liquid medium or on bacterial monolayers~\cite{Cas08}. 
These results might give clues to treat viral
infections of two-dimensional tissues: It has been shown that
space induces a remarkably strict clustering of the propagation in 
crops~\cite{Cou04}, in leaves~\cite{Tak07-2}, 
and in lawns of cells~\cite{Lee96}. 
Advances in therapy design ask for an improved understanding of the mechanisms causing
viral extinction, before they can be applied {\it in vivo}.

Models of viral evolution necessarily make simplifying assumptions, 
and real virus behavior often deviates substantially from their
predictions~\cite{Eig02}. 
Current quasispecies models implement high mutation rates that yield heterogeneous 
populations ---a property of
most viruses infecting plants~\cite{Gar01} and of many %{\it in vitro} systems of 
viral populations on cellular monolayers~\cite{Man06}. One common 
approximation is to assume that all new mutations have a deleterious effect 
on fitness, thus neglecting beneficial and neutral mutations. 
However, adaptation occurs frequently, and even at low population numbers. 
Classical quasispecies models assumed the existence of a unique 
master sequence of high fitness, while we currently know that 
a huge amount of different sequences yield 
phenotypes that perform equally well. 
When beneficial and neutral mutations are considered, extinction might
supervene through mechanisms other than error thresholds~\cite{Man10}.
The predictions yielded by models of viral evolution depend on whether
physical space is explicitly considered or not~\cite{Ran95}. In quasispecies models,
cell-to-cell transmission or local transport of viruses unavoidably results
in a lower value of the critical error threshold~\cite{Alt01,Toy05}.
Clusterization of viral types is responsible for the latter effect, as well as
for enhancing the coexistence of types in the quasispecies~\cite{Pas01,Agu08}. 

In this Letter we introduce a model representing the
propagation of infections on two dimensional arrays of cells. 
We incorporate two realistic features of infection propagation often
disregarded in quasispecies models: 
a non-negligible fraction of beneficial mutations 
and %the possibility that the host develops defenses against the pathogen. 
a host resistance to infection.
We show that with spare susceptible cells
%If there are no limits in the number of cells that can be infected
at each generation, the virus can escape host resistance by
increasing its progeny.
%the number of offspring per replication cycle. 
%However, when competition for cells within the quasispecies occurs, 
However, when viruses compete for infecting cells, increasing the progeny
does not pay and infection clearance occurs
at a finite value of the host resistance.
%to infection: an increase in the offspring numbers does not return
%the pathogen any further benefit.
In this case the model boils down to a multicomponent generalization of the
\emph{Domany-Kinzel} (DK) probabilistic cellular automaton~\cite{Dom84}
hence classifying viral extinction within the directed
percolation (DP) universality class, alongside with propagation
of immiscible fluids in porous media and other similar problems.

In the model, a viral population is described as an ensemble of (pheno)types,
each characterized by a replicative ability $r=0, \dots, R$
standing for the number of offspring able to infect cells 
produced by each type under replication. 
Deleterious or beneficial mutations decrease or increase (with probabilities
$p$ and $q$ respectively) offspring's replicative ability in one unit. 
%The offspring of a viral strain is affected by deleterious or beneficial
%mutations, decreasing or increasing its replicative ability in one unit,
%with deleterious and beneficial probability $p$ and $q$, respectively.
%Lethal mutations can hit the lowest class $r=1$ with probability $p$, and
Susceptible cells have a resistance to infection $\pi$, i.e.,
a viral particle can infect a cell with probability
$1-\pi$. For simplicity, we consider that the dynamics proceeds in discrete
generations. 

Assume first that the availability of susceptible cells is large
enough for every viral particle with $r>0$ to actually meet a cell to infect, 
no matter its replicative ability $r$. 
This is a mean-field dynamical description, where
infection occurs with a probability $1-\pi$. 
If $n_r(t)$ denotes the
number of viral particles of type $r$ at generation $t$,
for $1\le r <R$
%the dynamics
%of the population follows 
\begin{equation}
\begin{split}
n_r(t+1)=&\,(1-\pi) \left[ r (1-p-q) n_r(t)\right.\\
&+ \left.(r+1) p n_{r+1}(t) + (r-1) q n_{r-1}(t) \right], \\
n_R(t+1)=&\,(1-\pi) \left[ R (1-p) n_R(t) + (R-1) q n_{R-1}(t) \right].
\end{split}
\label{eq:meanfield1}
\end{equation}
Besides, there is a class $r=0$ which has lost its replicative ability
and is maintained by class $r=1$ through deleterious mutations, i.e.\
$n_0(t+1)=(1-\pi)pn_1(t)$. As initial condition we take $n_{r}(0)=\delta_{r,R}$.

An exact analytic solution exists for the case $q=0$,
which represents the worst situation for the survival of the virus:
if extinction does not occur for $q=0$, it will not occur
for $q>0$ either.
In the limit $t\to\infty$, the population of every class
grows at a fixed rate, $n_r(t+1)=\lambda n_r(t)$. System
\eqref{eq:meanfield1} then becomes an 
eigenvalue problem, $\lambda$ being
its largest eigenvalue. 
For $q=0$, 
the asymptotic growth rate is $\lambda = (1- \pi) R (1-p)$ and
the fraction of virus in class $r$ is
\begin{equation}
x_r\equiv\lim_{t\to\infty}\frac{n_r (t)}{\sum_{r=0}^Rn_r(t)}=
\binom{R}{r} p^{R-r}(1-p)^r.
\label{eq:stationary}
\end{equation}
Extinction occurs when $\lambda \le 1$,
%a condition fulfilled by all viral populations with a maximal
i.e., if the maximum progeny production $R \le R_c =
(1-\pi)^{-1} (1-p)^{-1}$. Although the larger the resistance to infection
of the cell $\pi$ the more demanding becomes this condition,
%on progeny production,
$R$ can always be increased beyond $R_c$ to avoid extinction.
%there is always a value of $R$ such that the asymptotic growth rate 
%of the population is above one.
If the number of susceptible
cells is unlimited and increasing $R$ always improves the chances of
survival of the pathogen, a runaway co-evolution between the virus and the 
host is to be expected.

The case $q=0$ is not representative of the situation when $q>0$ though.
The stationary distribution \eqref{eq:stationary} has a well defined
$R\to\infty$ limit, whereas the permanent improvement allowed when $q>0$
forbids a steady state in the same limit. As a matter of fact, it can be
shown that in that case the viral population grows superexponentially
\cite{Cue10}.
 
Consider now a restricted number of susceptible cells available. 
This occurs when cells have a spatial distribution and/or when the mobility  
of viruses is limited. To assess the consequences of this constraint, 
we study the dynamics proceeding on a triangular lattice with periodic 
boundary conditions. Even if infection starts off from a single
infected cell, the propagating front 
becomes asymptotically flat~\cite{Agu08}. 
Hence, we consider that cells form arrays in one spatial dimension, and infection
advances one row of cells per generation. 
Let $r_i(t)$ be the replicative ability of the individual occupying
site $i=1,\dots,L$ at generation $t$.

To account for the limited mobility of the viral particles, they are constrained 
to infect only neighboring cells. Hence, at generation $t+1$ site
$i$ will be occupied by one of the offspring of the individuals at sites
$i$ or $i+1$ at the parental generation $t$.
%Suppose that the replicative abilities of these individuals are
The probability that $r_i(t+1)=r$, given that
$r_i(t)=r_1$ and $r_{i+1}(t)=r_2$, is
\begin{equation}
\begin{split}
P(r |r_1,r_2) =& \pi^{r_1+r_2} \delta_{r,0} + \left( 1-\pi^{r_1+r_2} \right)
\frac{r_1p_{r_1,r}+r_2p_{r_2,r}}{r_1+r_2} \, ,
\end{split}
\label{PT2}
\end{equation}
where $p_{r,r-1}=p$ for $r \le R$, $p_{r,r+1}=q$ and $p_{r,r}=1-p-q$ for $r<R$,
$p_{R,R+1}=0$, and $p_{R,R}=1-p$. We take $r_i(0)=R$ for all $i=1,
\dots,L$ when the infection begins. According to~\eqref{PT2}, the probability that 
the cell resists the viral attack 
is given by $\pi^{r_1+r_2}$, i.e. the probability 
that infection fails after $r_1+r_2$ independent trials. 
With the complementary probability $1-\pi^{r_1+r_2}$ infection occurs, and
one individual randomly drawn among the $r_1+r_2$ possible occupies the
site. Its replicative ability may change due to mutations as specified.
The limit case $\pi \to 1$ describes an immune host, whereas 
in the limit $\pi \to 0$ the host plays no role in
the progress of infection. 
Two relevant quantities that characterize the dynamical state of the system are
the spatial density $\rho(t)$ of virus with $r \ge 1$
at time $t$ and the average replicative ability of the population
$\bar r(t) = (\rho(t) L)^{-1} \sum_{i=1}^L r_i(t)$. 
A systematic study of these two quantities in the limit $t \to \infty$ as
a function of $p$ and $\pi$ portrays the asymptotic behavior of the model.

\begin{figure}
\vspace*{-2mm}\begin{center}
\includegraphics[width=3.3in,clip=]{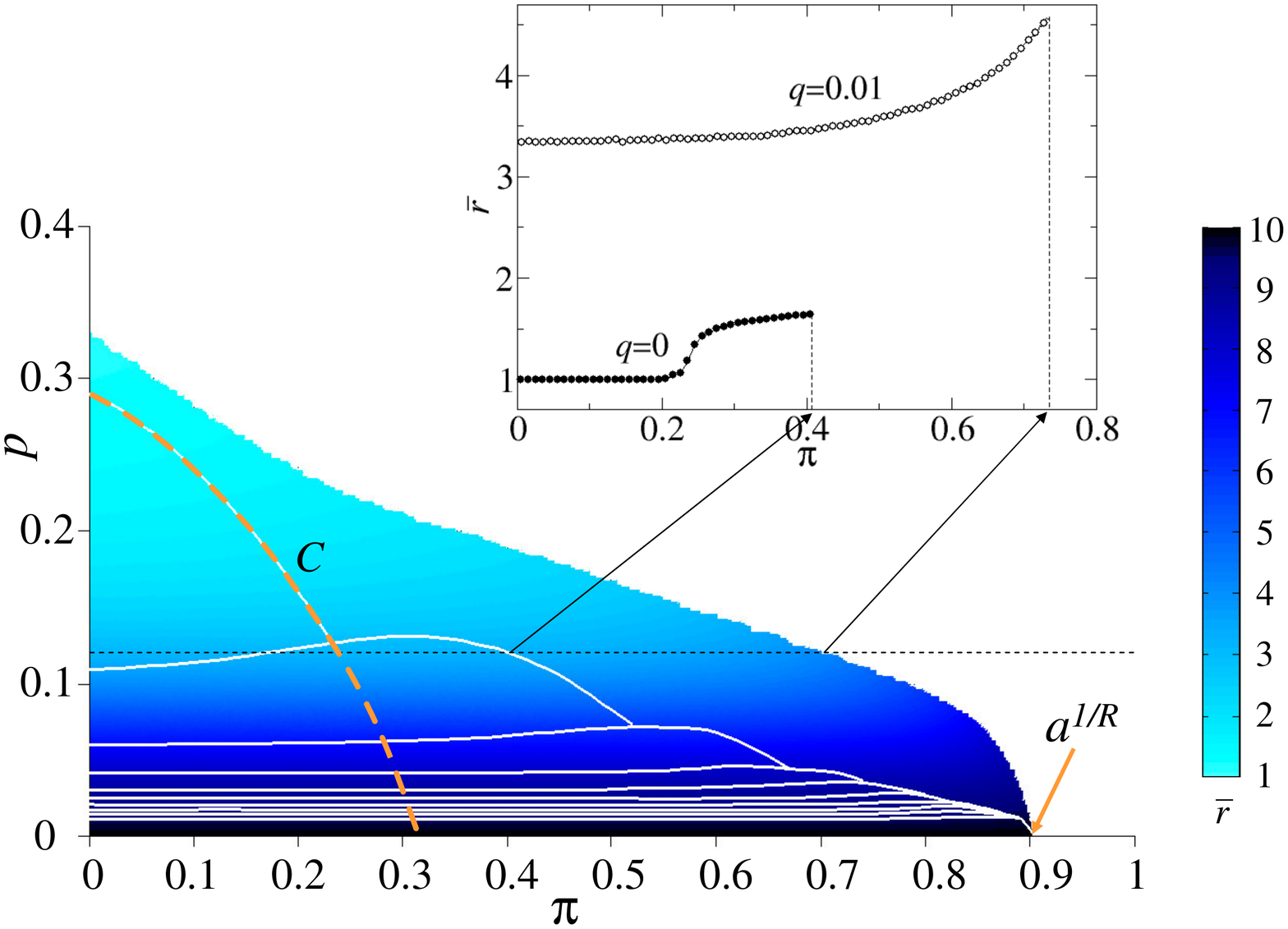}
\end{center}
\vspace*{-8mm}\caption{
(Color online) Phase space of the model with $R=10$
and two choices for $q$:
$q=0$, where the highest$-r$ classes are sequentially lost 
as $p$ increases through transitions analogous to the error threshold of
Eigen's quasispecies model (white lines),
and $q=0.01$, where only global extinction 
of the population is possible.
%In these examples, $R=10$.
%White lines correspond to 
%successive losses of the highest replicative class present ($r$
%decreasing from bottom to top) when $q=0$.
Colors code for the average
replicative ability $\bar r$
%as a function of $p$ and $\pi$
when $q=0.01$. In the white domain, infection is cleared
in finite time. The top extinction occurring for
$q=0$ can be obtained from that of the DK cellular automaton (curve $C$). 
The inset shows $\bar r$ for $p=0.11$ and two values of $q$. 
Data have been obtained with a system
of size $L=10^4$ and averaging over $5\times 10^3$ time steps after a
transient of $2 \times 10^4$ generations. 
}
\label{F-phasespace}
\end{figure}

In the absence of beneficial mutations ($q=0$),
the class with the highest $r$ present disappears when $p$ or $\pi$ increase beyond
a threshold value. These transitions are somehow analogous to the error threshold
described for simple models of quasispecies~\cite{Eig71}, where the master
sequence disappears and the population turns into a cloud of low-fitness
replicating mutants. 
For constant resistance to infection $\pi$,
there is a hierarchy of domains in the phase space where high$-r$ classes
are sequentially lost as $p$ increases 
(see Fig.~\ref{F-phasespace}). For increasing
$\pi$, the largest $r-$value present at global extinction increases.

There are two features of Fig.~\ref{F-phasespace} for $q=0$ which are
amenable to further analysis. The first one is the last error threshold,
when class $r=1$ disappears. There is a mapping
of this transition to the DK cellular automaton.
When only classes $r=1$ and $r=0$ are present,
sites are either replicative (occupied) or
non-replicative (empty), yielding 
probabilities \eqref{PT2} $P(1|0,0)=0$, 
$p_1\equiv P(1|1,0)=P(1|0,1)=(1-\pi)(1-p)$,
and $p_2\equiv P(1|1,1)=(1-\pi^2)(1-p)$, which define a DK automaton.
Curve $C$ in
Fig.~\ref{F-phasespace} coincides with the DK transition cast in the
variables ($p$, $\pi$) describing our model.
%The case $p=0$, corresponding to bond percolation, 
%is irrelevant for any $R > 1$ because it
%is preempted by the extinction transitions of higher
%$r$ classes. However, in the
This curve meets the $\pi=0$ axis
at $p_c=0.294510(6)$, 
%(see Fig.~\ref{F-phasespace}),
a value derived
from the known site percolation critical point
\cite{Hin00}.

The second feature obtained from a mapping with the
DK model is the extinction threshold of the highest class $r=R$ at $p=0$.
Since the initial state only contains individuals of this type, no
other types will ever appear because of the lack of
deleterious mutations. 
Hence, sites can be either occupied or empty, just as in the DK model. 
Transition probabilities 
correspond to the bond percolation case, from which it follows that the
threshold occurs at $\pi_c=a^{1/R}$, with $a=0.3552998(9)$. Only in the limit 
$R \to \infty$ would $\pi_c \to 1$, taking a lower value for any
other combination of parameters.  

For $p \ne 0$ and $q>0$, no matter how small, 
all replicative types are simultaneously present in the infective phase 
(see Fig.~\ref{F-phasespace}). For a fixed value of
%the resistance to infection
$\pi$, lower $r$ classes get
more populated upon increasing $p$, hence the average replicative ability
decreases. The behavior as $\pi$ increases at constant $p$ is of a different nature.
As $\pi$ increases, infected cells become sparser. In the absence of neighbors,
the probability that a cell at generation $t+1$ is infected by offspring 
of an individual of class $r$ at generation $t$ is proportional to $(1-\pi^r)$. 
This quantity penalizes more severely low$-r$ classes as $\pi$ increases, such 
that high$-r$ classes receive a relative advantage and $\bar r$ increases 
as $\pi_c$ is approached. 
At the critical value $\pi_c$ the quasispecies eventually collapses because
propagation is prevented by host defenses (see the inset of 
Fig.~\ref{F-phasespace}). 

\begin{figure}[!t]
\vspace*{-7mm}\begin{center}
\includegraphics[height=3.5in,angle=90,clip=]{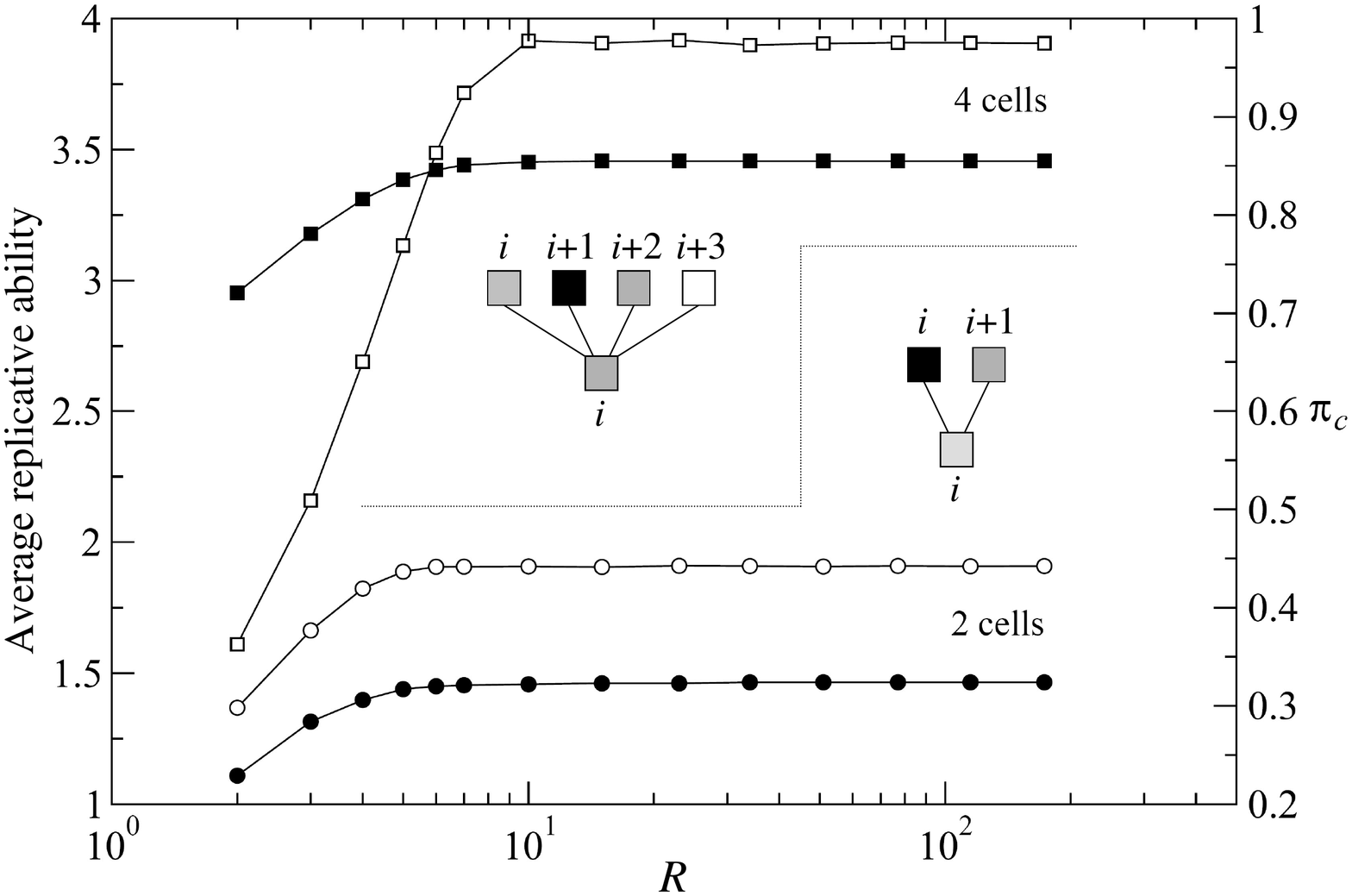}
\end{center}
\vspace*{-8mm}\caption{
When the number of susceptible cells available is limited, 
the average replicative ability $\bar r$ (open symbols,
left $y-$axis) and the critical value of the resistance to infection $\pi_c$ 
(solid symbols, right $y-$axis) saturate to finite values as 
$R$ increases. In these curves, $p=0.2$, $q=0.01$,
and averages over 50 independent realizations for each value of $\pi_c$ have
been performed. Mean-square deviations are smaller than symbol size.}
\label{Sat}
\end{figure}

There is another relevant difference between the two models.
As we have seen, when cells are in excess
any host resistance $\pi$ can be circumvented by increasing $R$.
%there is a value of $R$ for each $\pi$ that would
%guarantee propagation of infection.
In contrast, when viral classes have to 
compete for susceptible cells, there is a critical value of $\pi$ above
which this is no longer true.
%the propagation of the viral infection comes to a halt. 
The reason (illustrated by Fig.~\ref{Sat}) is that 
the average replicative ability $\bar r$ quickly saturates to a constant
value upon increasing $R$. In other words,
when viruses struggle for infecting a limited number of cells,
increasing $R$ does not provide any additional advantage.
%To illustrate this statement, we have analyzed how
%the average replicative ability $\bar r$ varies upon increasing $R$ at 
%the value of $\pi$ where extinction supervenes (Fig.~\ref{Sat}).
%We observe that $\bar r$ quickly saturates to a constant value. 
In order to show that this effect results from the limitation in the number of
cells available, we have modified the model to consider 
a case where four cells are available for each infection event. This
amounts to increasing the competition between types, such that low$-r$ classes
are at a disadvantage with respect to the previous situation and 
$\bar r$ increases~\cite{Agu08} (see Fig.~\ref{Sat}).
%as we indeed observe.
As a side effect, the critical value of the cell resistance,
$\pi_c$, at which extinction occurs also saturates to a value below
$1$ as $R\to\infty$. A runaway co-evolution is therefore
excluded in this case: if the host cell is able to increase its defenses beyond
$\pi_c$ the infection will not progress. 

\begin{figure}[!t]
\begin{center}
\includegraphics[width=2.8in,clip=]{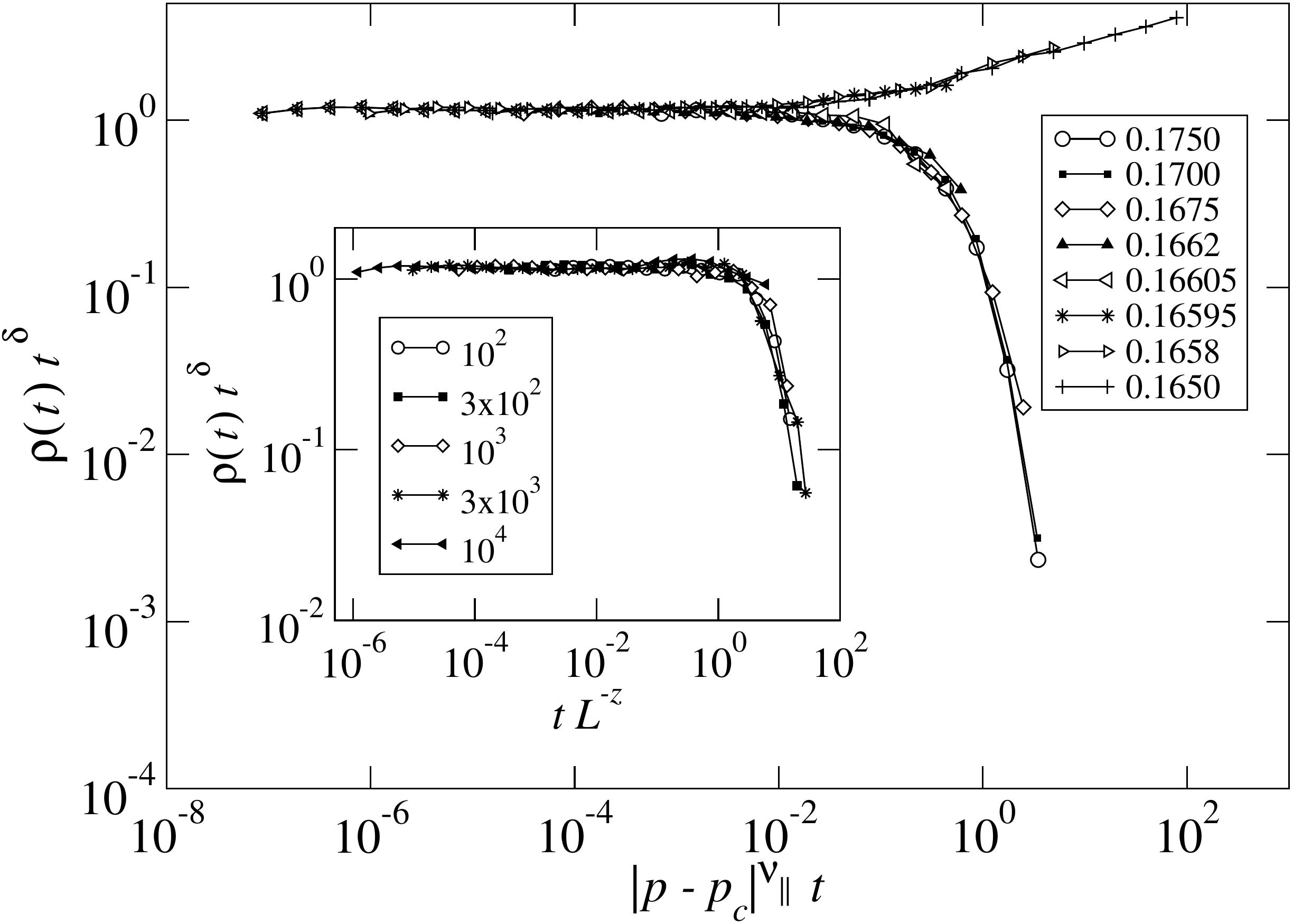}
\end{center}
\vspace*{-3mm}
\caption{
Data collapse of curves $\rho(t)$ according to the critical exponents
known to characterize (1+1) directed percolation,
which are $\delta=0.159464(6)$, $\nu_{\parallel}=1.733847(6)$,
and $z=1.580745(10)$~\cite{Hin00}. Parameters in these simulations 
are $R=2$, $q=0.01$, $\pi=0.3$. The system size is
$L=10^4$ for the main plot, where the legend shows the values of $p$ for
each curve. For this choice the transition occurs at $p_c=0.16600(3)$.
Legend in the inset gives the values of the system size $L$. Data have been
averaged over $50$ independent realizations and up to $10^8$ generations
for each curve.
}
\label{F-collapse}
\end{figure}

When the number of susceptible cells is limited, the transitions to extinction
%both of the whole population or of the class with the highest replicative
%ability present ($q=0$),
belong to the universality class of DP for any $q\ge 0$.
For $q=0$ we have rigorously proven this for the extinction of the $r=1$ class 
as well as when $p\to 0$, by mapping the model to a 
DK cellular automaton, known to belong to the DP class~\cite{Hin00}.
Besides, our model is defined on a set of
local rules, has no special symmetries or disorder, and exhibits a continuous
transition to a unique absorbing state (total extinction).
Therefore, according to the DP conjecture it should belong
to the DP universality class. An empirical proof is provided
in Fig.~\ref{F-collapse}, where $\rho(t)$ is
appropriately rescaled according to the exponents of DP. 
The transition point $p_c$ is 
numerically determined as the value of $p$ yielding the best collapse.
Finally, the study of the variation of 
$\rho(t)$ at $p_c$ with the system size $L$ is characterized by the dynamic 
exponent $z=1.580745(10)$. This analysis has been repeated for the
transition where $r=2$ disappears at $q=0$, with analogous results. 

Formally, the probabilistic model presented here
%, described by the probabilistic rules $P(r_3|r_1,r_2)$,
defines
a multicomponent generalization of the DK cellular automaton. Our
model only focuses in a limited region of its parameters in which
an active-inactive transition ($q>0$) or several ($q=0$) occur with
the criticality of directed percolation. Exploring in depth the
whole phase space of this model in search for deviations of this
behavior is an interesting open question left out by the present study.

Our model could be modified to better mimic other biological situations. 
In many plants, different individuals display variable degrees of resistance to viral
infection~\cite{Kan05}, a case that could be implemented by introducing a host-dependent
$\pi$. This would introduce a form of quenched spatial 
disorder that may lead to universality classes other than
DP~\cite{Hin00}. Analogous changes would be
effected by temporally quenched disorder, a situation holding, e.g.,
when susceptibility to infection depends on the age of the host. 
An example of the latter is systemic acquired resistance, a
durable form of immunization observed in plants~\cite{Dur04}. Spatially
quenched disorder could change the properties of viral extinction to
those of dynamic percolation~\cite{Gra83}. 
Current knowledge on the properties of percolation phenomena in a
variety of different situations might inspire unprecedented strategies
to stop infection propagation in those different environments.

The hindrance to the viability of a virus brought about by a
limited avaliability of cells has 
%It is to be expected that the viability of a virus would be 
%compromised when the number of available cells is reduced.
%There is
some empirical support.
%for this prediction.
While all plant
viruses infect the tissue that carries nutrients through the vascular
system
%of the plant
(phloem), where mobility is high and
%the number of
susceptible cells are abundant,
%is relatively large, not all viruses are
some viruses (e.g.\ luteoviruses and some geminiviruses~\cite{Kni07}) are not
able to infect leaves, where mobility is limited to cell-to-cell
transmission and fewer uninfected cells are available per infective cycle.
%Such is the case of luteoviruses and some geminiviruses
%whose infections affect the plant phloem~\cite{Kni07} but never reach leaves.
The reverse situation is as yet unknown. 

The authors acknowledge conversations with E.~L\'azaro, and the
support of the Spanish MICINN 
under projects FIS2008-05273 and MOSAICO, and of Comunidad de Madrid (Spain)
under project MODELICO-CM.

%\bibliography{virus}
%

\end{document}